**Short Paper**

# A Game-Based Learning Application to Help Learners to Practice Mathematical Patterns and Structures

Adrian S. Lozano
College of Computing Studies, Mexico Campus, Don Honorio Ventura State University

Reister Justine B. Canlas
College of Computing Studies, Mexico Campus, Don Honorio Ventura State University

Kimberly M. Coronel
College of Computing Studies, Mexico Campus, Don Honorio Ventura State University

Justin M. Canlas
College of Computing Studies, Mexico Campus, Don Honorio Ventura State University

Jerico G. Duya
College of Computing Studies, Mexico Campus, Don Honorio Ventura State University

Regina C. Macapagal
College of Computing Studies, Mexico Campus, Don Honorio Ventura State University

Ericson M. Dungca
College of Computing Studies, Mexico Campus, Don Honorio Ventura State University

John Paul P. Miranda
College of Computing Studies, Mexico Campus, Don Honorio Ventura State University
jppmiranda@dhvsu.edu.ph
(corresponding author)

Recommended citation:

**Abstract**

*Purpose* – The purpose of this study is to develop a game-based mobile application to help learners practice mathematical patterns and structures.

*Method* – The study followed a mixed-method research design and prototyping methodology to guide the study in developing the mobile application. An instrument based on the Octalysis framework was developed as an evaluation tool for the study.

*Results* – The study developed a mobile application based on the Octalysis framework. The application has fully achieved all its intended features based on the rating provided by the students and IT experts.

*Conclusion* – The study successfully developed a mobile learning application for mathematical patterns and structures. By incorporating GBL principles and the Octalysis framework, the app achieved its intended features and received positive evaluations from students and IT experts. This highlights the potential of the app in promoting mathematical learning.

*Recommendations* – This study recommends that the application be further enhanced to include other topics. Incorporating other game-based principles and approaches like timed questions and the difficulty level is also worth pursuing. Actual testing for end-users is also needed to verify the application's effectiveness.

*Practical Implications* – Successful development of a game-based mobile app for practicing mathematical patterns and structures can transform education technology by engaging learners and enhancing their experience. This study provides valuable insights for future researchers developing similar applications, highlighting the potential to revolutionize traditional approaches and create an interactive learning environment for improving mathematical abilities.

## INTRODUCTION

Mathematical pattern is the heart of mathematics (Mulligan et al., 2010; Platas, n.d.). Since mathematics is all about numbers, it involves studying different patterns in the universe (Mulligan et al., 2010). Examples of these patterns include number sequences, logic, imagery, word patterns, etc. It is also a regularity found in how numbers or other mathematical objects behave or relate to one another (Mason et al., 2009; Mulligan et al., 2010). Patterns can be described and analyzed using mathematical concepts such as functions, sequences, and algebraic expressions. In contrast, the pattern can be described and analyzed using mathematical concepts such as functions, sequences, and algebraic expressions. Furthermore, structures in a pattern usually involve arithmetic operations. An example of these is arithmetic sequences and algebraic expressions. Understanding mathematical patterns and structures is also a way for students to learn about the world around them (Mason et al., 2009). It allows the understanding and description of how the world revolves. Likewise, students need to understand this area of mathematics due to its wide-ranging applications in many fields (e.g., applied science, engineering, and economics). Skills acquired in understanding various concepts under mathematical pattern and structures is crucial for solving problems, making predictions, and making connections to various areas of mathematics. Furthermore, patterns and structures can also be used to create complex systems and data-driven decisions. This area is also used in various mathematics-related areas, from predicting the potential spread of diseases and planetary movement to the stock market movement. Regarding technology, pattern and structures revolves around creating and designing efficient systems not limited to transportation and manufacturing. A student who fully understands this area often develops greater critical and logical thinking and problem-solving skills (Mulligan & Mitchelmore, 2009).

In the Philippines, mathematical patterns and structure are taught among Grade 10 students. The country is ranked second lowest among the 79 participating countries according to the latest OECD's Programme for International Student Assessment (Paris, 2019) and ranked lowest among the 58 countries participating in the recent Trends in International Mathematics and Science Study reports (CNN Philippines, 2020). This study was conceived to help learners and provide an alternative means for learners to practice and learn mathematical patterns and structures using a game-based approach. A student can learn the topic more interactively and interestingly by developing an application for the said topic. Using a mobile application intended for a specific topic, students can actively participate in the learning process, which may improve comprehension and memory, unlike traditional learning methods where students are only expected to listen and take notes. In addition, previous studies have shown that technology improves student engagement, motivation, and overall learning outcomes (e.g., Simsek, 2016; Dele-Ajayi et al., 2019; Gil-Doménech & Berbegal-Mirabent, 2019; Serrano, 2019). A topic can become more enjoyable and accessible by incorporating gamification elements.

## LITERATURE REVIEW

### Game-based Learning for Mathematics

Game-based learning (GBL) is a teaching approach incorporating educational games into the learning process (Pho & Dinscore, 2015). It is considered an active learning technique to enhance student learning into which existing and innovative games are often incorporated (Whitton, 2012). GBL can be an effective way to engage and motivate students, particularly when it comes to learning mathematics (Beserra et al., 2017; Cózar-Gutiérrez & Sáez-López, 2016; Dele-Ajayi et al., 2019; Gil-Doménech & Berbegal-Mirabent, 2019; Serrano, 2019). Furthermore, GBL allows students to learn without realizing it (Tokac et al., 2019). This is because GBL is an approach that is often and most of the time crafted based on specific learning objectives (Beşaltı & Kul, 2021; Bringula et al., 2018; Simsek, 2016; Tao et al., 2018).

There are several methodologies and key components under GBL, including serious games, gamification, and simulation-based learning. In the context of learning mathematics using serious games, users are taught and trained on specific mathematical skills. In gamification, typical game elements (i.e., scoreboard, mechanics, points, badges etc.) are incorporated to motivate and engage the student to achieve a specific learning objective. While in simulation-based learning, creating a realistic learning environment is needed. Oftentimes, simulations are created to provide hands-on experience to a wide range of subjects, although they may be costly to develop.

There are many GBL applications for mathematics, ranging from simple arithmetic games to more complex problem-solving and critical-thinking games. For example, Dele-Ayjayi et al. (2019) and Gil-Doménech and Berbegal-Mirabent (2019) indicate that GBL help stimulates students' interest and engagement in mathematics (Simsek, 2016). Similar observation was found by Beserra et al in 2017 and Cózar-Gutiérrez and Sáez-López in 2016. In addition to these, Garcia and Mangaba (2017) stipulated in their study that GBL is often used for curricular knowledge.  In addition to these, several studies have also pointed out that GBL, together with technology, can further support student learning in various areas (e.g., Behnamnia et al., 2022; Beşaltı & Kul, 2021; Bringula et al., 2018; Garcia, 2020; Godoy Jr., 2020). Other studies have also indicated that GBL with technology creates a positive effect on students as well as on achieving learning outcomes (Cahyana et al., 2017; Chao et al., 2018). Therefore, GBL applications for mathematics can be a fun and engaging way for students to learn and practice math skills and can be used in various educational settings, from the classroom to online learning environments. Incorporating and leveraging GBL's game design principles can provide an alternative approach that may engage and motivate students to learn mathematics.

## The Octalysis Framework

The Octalysis framework is a gamification framework developed by Yu-kai Chou in 2015. This framework comprises eight core drives: epic meaning and calling, development and accomplishment, empowerment of creativity and feedback, ownership and possession, social influence and relatedness, unpredictability and curiosity, and loss and avoidance. Chou (2015) defined the first driver, epic meaning and calling, as an aspect to which players believed that they chose to do something. The second driver, development and accomplishment, focuses on the user's drive to improve, progress, and overcome challenges. These things are often seen in leadership boards and badges, and points. The third driver, empowerment of creativity and feedback, was defined by Chou as something that makes a player engage in something that can be brought out their creativity. The fourth driver, ownership and possession, motivates players because they feel ownership of something and the urge to improve what they own. The fifth driver, social influence and relatedness, is about companionship, social responses, competition, and envy. The sixth driver is scarcity and impatience. This driver is focused on making a player want to own something that is only achievable sometimes. The seventh driver is unpredictability and curiosity. This driver is about the randomness of things – rendering the players to guess what will likely happen next. The last eight drivers are loss and avoidance. This driver is based on the concept that something bad or negative might happen, resulting in potential loss; therefore, avoidance is necessary.

In this study, the Octalyis framework was selected primarily for its comprehensive use in gamification. It is particularly structured for designing gamified systems. The said framework has been previously used in various contexts, not only in education but also in other domains. This makes the framework more reliable and established and can be applied in various contexts like educational learning. In addition, the framework provides a simple, concise, and clear guide for developers in creating GBL software. Compared to existing frameworks like Self-determination Theory (SDT), which focuses on psychological needs, competence, relatedness for motivation and engagement, and Player Experience (PX), they only focus on users' emotional and social experiences to create engaging gameplay. The Octalysis framework is more practical and extensive in providing a guide to making game-based software.

### Research Objectives

The study aimed to develop a game-based mobile learning application for mathematical patterns and structures.

Specifically, the study intended to achieve the following:

1. Develop a mobile application about mathematical patterns and their structure;
2. Employ existing GBL principles and approaches to the proposed application; and
3. Evaluate the developed application using the Octalysis principles.

## METHODOLOGY

### Research Design and Software Development Model

The study followed a mixed-method approach consisting of a short survey and semi-structured interviews to select respondents. The short survey used in this study is developed based on Octalysis framework which consisted of 13 questions. For the interviews, respondents were asked on how to further improve the prototype until it reaches the intended functionalities based on the aforementioned framework. This process is in line with the software development process. This was followed to identify how the application should look and what specific areas should be considered in developing the application. The prototyping methodology was specifically followed to achieve the specific features based on the topic and the requirement of potential application end-users. The prototype model is ideal for developing an application whose requirements and specific details are known after a period of time. It was also ideal for this study due to its iterative and trial-and-error process between developers and end-users (Lewis, 2019).

## Software Development Process

This study followed a six-phase prototyping model, as seen in Figure 1, requirement gathering, quick design, building prototype, customer evaluation, refining prototype, and engineering product. In the requirement gathering, this study carefully reviewed the existing literature and observed how to create a game-based application for mathematical patterns and structure and what type of framework should be applied. When the requirements were known, preliminary designs of the mobile application based on the data gathered in requirement gathering and initial drawing outlines were crafted and transformed into digital storyboards (i.e., interface designs). Afterward, initial prototypes were created and evaluated (i.e., in the form of impressions and initial feedback) by potential users were acquired. These feedbacks were then taken into consideration to refine the mobile application further and proceeded to another quick designing process until the initial application was completed and finally evaluated by intended users and IT experts.

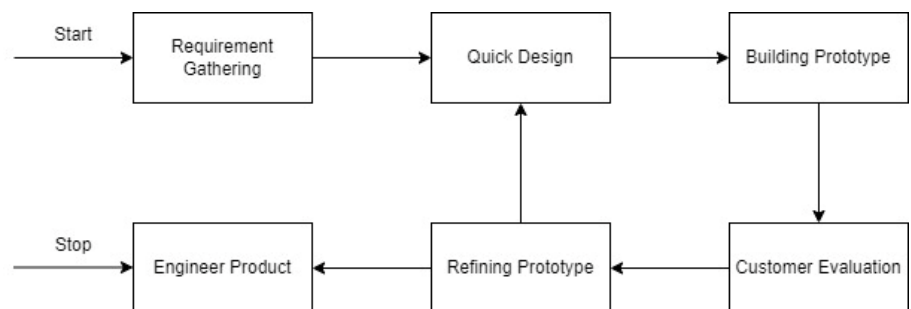

*Figure 1.* Prototype Model (Shah & Dhiman, 2016).

## Study Respondents, Sampling Technique, and Sample Size

There are two types of respondents for this study. The first group of respondents was IT-related experts with at least three years of experience in the industry. These IT-related experts have previously developed educational software on various platforms. These experts were chosen due to their expertise in software development. In contrast, the second group was students from a senior high school in Mexico, Pampanga, Philippines. Initial consent was asked for them to participate in the study. A purposive sampling technique was utilized for this study. A total of three information technology (IT) experts accepted to be part of this study, and a total of 60 grade 10 students voluntarily accepted to be respondents for this study.

## Research Instrument and Data Analysis

In achieving the third objective of this study, a questionnaire was developed, and initially, content validated by three experts (i.e., educators with a background in educational technology) instrument was used. The instrument consists of 13 questions which were based on the Octalysis framework, which comprised of epic meaning and calling, development and accomplishment, empowerment of creativity and feedback, ownership and possession, social influence and relatedness, scarcity and impatience, unpredictability and curiosity, and loss and avoidance (Figure 2). Each item was rated using a 4-point Likert scale (i.e., fully achieved, achieved, partially achieved, and not achieved). Descriptive statistics of the evaluation result, such as frequency, mean, and interpretation, were reported.

Table 1. Range and interpretation of the results

| Scale | Range | Verbal Interpretation | Short Description |
|-------|-------|----------------------|-------------------|
| 1 | 1.00 – 1.75 | not achieved | 25% of the intended functionality are working |
| 2 | 1.76 – 2.50 | partially achieved | 26 - 50% of the intended functionality are working |
| 3 | 2.51 – 3.25 | achieved | 51 - 75% of the intended functionality are working |
| 4 | 3.26 - 4 | fully achieved | 76 – 100%25% of the intended functionality are working |

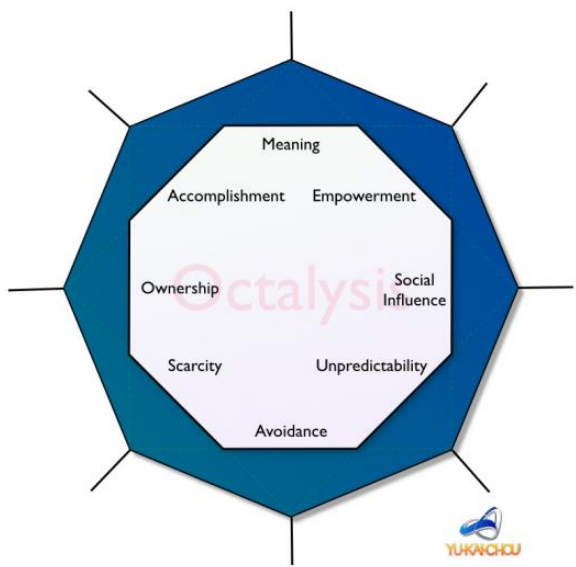

*Figure 2. Octalysis framework (Chou, 2015).*

## RESULTS AND DISCUSSION

The application will initially ask users to register by providing a name. This feature is needed to allow the application to save whose profile to load and retrieve its historical data. An initial avatar will be given to players (Figure 3). Figure 3 further shows the starting page of the application – named Home UI. The Home UI allows users to edit game settings such as music and volume. The game tutorial for beginners, current quests, and the user profile can also be found in this UI (Figure 3). Figure 3 also shows the game shop UI of the application. The user can purchase new avatar designs using application points in this UI. Figure 4, on the other hand, shows the tutorials provided by the applications and the actual game within the application. The study used the concepts of snakes and ladders paired with questions and answers about mathematical patterns and structure. A dice can be rolled every time the user wants to proceed with the game and continue answering the questions. In addition, Figure 5 shows the different feedback and notifications provided by the application. There are four types of feedback, i.e., "Good job!" for correct answers, "Ooppss!" for a wrong answer, "Victory" for clearing the game, and "Game Over" when lifelines were depleted.

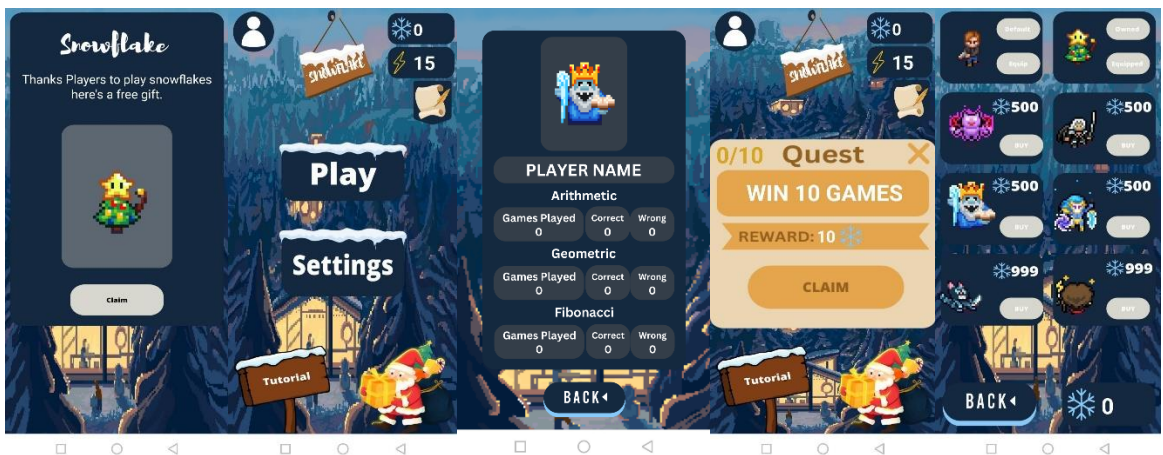

*Figure 3*. Free avatar, home user interface, profile, daily quests, and game shop

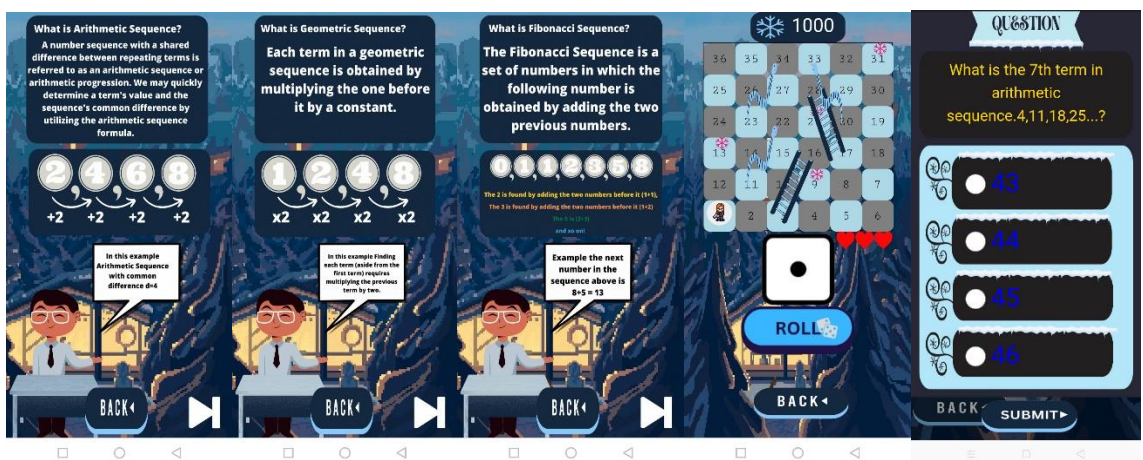

*Figure 4*. Tutorials and actual game within the application

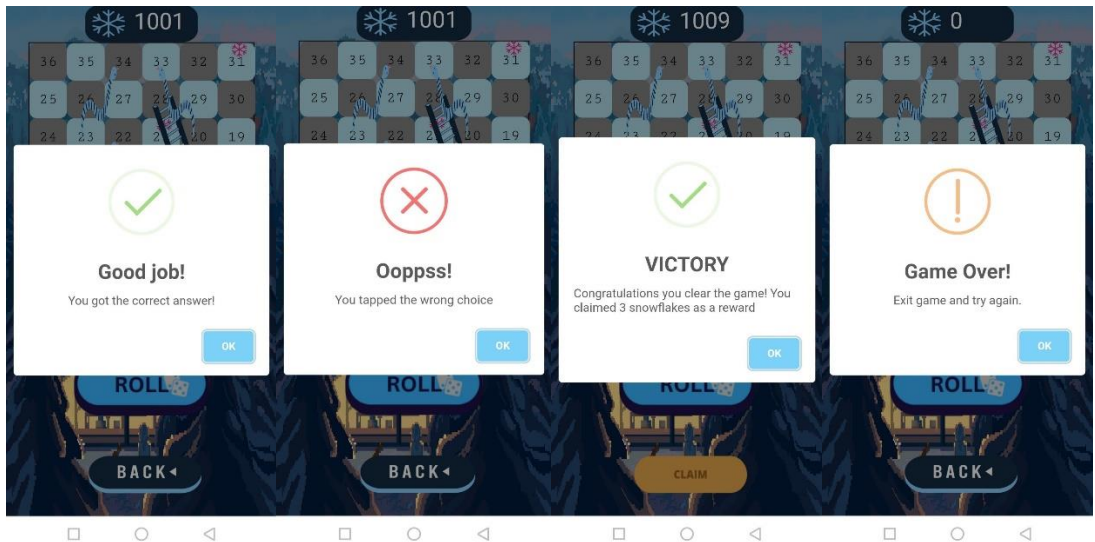

*Figure 5.* Game feedback and notifications

In this study, the Octalysis framework was a basis for GBL. The epic meaning and calling were applied by creating a free avatar that end-users (i.e., players) can use. Development and accomplishment were applied by adding the ability to save end-users progress in the game and check their existing points, scores, and current achievements (i.e., rewards). Creativity and feedback were empowered by allowing end-users to obtain points and rewards throughout the game. Extra points are also given for every completion of daily quests and continued playing. Ownership and possession were applied by having the ability to purchase additional avatars. Social influence and relatedness were applied by making the game similar to Snakes and Ladders. Scarcity and impatience were applied by limiting the ability of the end-user to play the game. Unpredictability and curiosity were applied by adding total randomness to the games and questions and answers. Loss and avoidance were applied by ensuring end-users could earn a lifeline to continue playing the games.

Table 2 shows the software evaluation of the prototype. Based on Table 2, the application has fully achieved all the criteria based on the Octalysis framework. Regarding epic meaning and calling, the application received a mean of 3.71 from students and 4.00 from IT experts. For development and accomplishment, it received a mean of 3.62 from students and 3.66 from IT experts. Some usability and technical issues concerning the saving feature of the prototype were encountered during the testing of the application, as mentioned by both evaluators.

For empowerment of creativity and feedback, it received a mean of 3.59 from students and 4.00 from IT experts. For ownership and possession, it received a mean of 3.62 from students and 4.00 from IT experts. For social influence and relatedness, it received a mean of 3.68 from students and 4.00 from IT experts. It received a mean of 3.69 from students and 4.00 from IT experts for scarcity and impatience. For unpredictability and curiosity, it received a mean of 3.67 from students and 4.00 from IT experts. In summary, students and experts indicated that the applications were working most of the time properly.

Table 2. Summary of evaluation results

| Question | Student mean (n = 60) | Interpretation | IT experts mean (n = 3) | Interpretation |
|---|---|---|---|---|
| Epic meaning and calling | | | | |
| I was able to answer the questions posed in the game until I reached the finish tile. | 3.72 | Fully Achieved | 4.00 | Fully Achieved |
| I was able to use a free avatar in playing the game. | 3.70 | Fully Achieved | 4.00 | Fully Achieved |
| Development and accomplishment | | | | |
| I was able to gain points, scores, rewards and able to check user profile. | 3.70 | Fully Achieved | 4.00 | Fully Achieved |
| I was able to save my progress every time. | 3.62 | Fully Achieved | 3.66 | Fully Achieved |
| Empowerment of creativity and feedback | | | | |
| I was able to earn more points and rewards through continues playing. | 3.67 | Fully Achieved | 4.00 | Fully Achieved |
| I was able to get an extra reward for every completion of a daily quest. | 3.50 | Fully Achieved | 4.00 | Fully Achieved |
| Ownership and possession | | | | |
| I was able to exchange my points into a new avatar design at the game shop. | 3.62 | Fully Achieved | 4.00 | Fully Achieved |
| Social influence and relatedness | | | | |
| I was able to solve the questions through the assistance of the game tutorial. | 3.65 | Fully Achieved | 4.00 | Fully Achieved |
| I was reminded of the gameplay of the snake and ladder while learning. | 3.70 | Fully Achieved | 4.00 | Fully Achieved |

Table 2. Summary of evaluation results (cont.)

| Question | Student mean (n = 60) | Interpretation | IT experts mean (n = 3) | Interpretation |
|---|---|---|---|---|
| Unpredictability and curiosity | | | | |
| I was able to re-play the game time after time depending upon the available energy. | 3.67 | Fully Achieved | 4.00 | Fully Achieved |
| I was able to obtain the most expensive avatar, enabling me to keep playing and earn more rewards. | 3.70 | Fully Achieved | 4.00 | Fully Achieved |
| Loss and avoidance | | | | |
| I was able to think about the questions I need to answer every time I roll the dice. | 3.75 | Fully Achieved | 4.00 | Fully Achieved |
| I was able to ensure that my answers to each question are correct; thus, able to avoid losing a lifeline. | 3.67 | Fully Achieved | 4.00 | Fully Achieved |

## CONCLUSIONS AND RECOMMENDATIONS

The study developed a mobile learning application for mathematical patterns and structures. The integration of GBL principles and approaches based on the Octalysis framework was also achieved based on the initial evaluation of the application. The applications have achieved all their intended features. The end-users and IT experts have evaluated it as fully achieved based on the eight criteria of the aforementioned framework.

This study recommends that the application be further enhanced to include other topics within the application. Incorporating other game-based principles and approaches like timed questions and the difficulty level is also worth pursuing. Actual and experimental testing is also needed to determine if the application was useful in helping the student learn and enhance their knowledge and skills about mathematical patterns and structures.

## PRACTICAL IMPLICATION

The successful development and implementation of a game-based mobile application for practicing mathematical patterns and structures can have significant implications for educational technology. The findings of this study suggest that incorporating gamification elements, such as the Octalysis framework, is achievable and might be able to engage learners and enhance their learning experience. In addition, this kind of study and software developments have the potential to transform conventional approaches to difficult subjects or topics like the one in this study (i.e., mathematical patterns and structure). By developing this kind of educational software technologies, it can aid in providing a fun and interactive environment for students to learn and improve their mathematical ability. In research, this study provided valuable insights for future researchers that might want to develop similar educational game-based applications.

## ACKNOWLEDGEMENT

No funding was received for this study.

## DECLARATIONS
### Conflict of Interest

The authors declare no conflict of interest in this study.

### Informed Consent

Individual consent based on the data privacy laws in the Philippines was asked of the study respondents.

### Ethics Approval

This study involved human participation through the distribution of survey forms to students and IT experts, as well as the conduct of interview sessions. Measures were taken to ensure the well-being, privacy, and integrity of participants, adhering to relevant guidelines, regulations, and standards in accordance to the Philippine Data Privacy Act and the Philippine Health Research Ethics Board.

## Author's Biography

**Adrian S. Lozano, Reister Justine B. Canlas, Kimberly M. Coronel, Justin M. Canlas, Jerico G. Duya, Regina C. Macapagal,** and **Ericson M. Dungca** are student researchers in the field of Information Technology at the College of Computing Studies of Don Honorio Ventura State University, Mexico Campus.

**John Paul P. Miranda** is an assistant professor of information technology at Don Honorio Ventura State University. He serves as the program head for international linkages and partnerships in the Office of International Partnerships and Programs, overseeing collaborations and partnerships with international institutions and organizations. Additionally, he holds the role of Secretariat for the University Scholarship Grant Committee at the university.